\documentclass[conference]{IEEEtran}
\IEEEoverridecommandlockouts
\pdfoutput=1
\usepackage{cite}
\makeatletter
\newcommand{\removelatexerror}{\let\@latex@error\@gobble}
\makeatother
\usepackage{amsmath,amssymb,amsfonts}
\usepackage{graphicx}
\usepackage{subfigure}
\usepackage{textcomp}
\usepackage{xcolor}

\usepackage{algorithmic}
\usepackage{bm}
\usepackage{diagbox}
\usepackage{booktabs}
\usepackage{url}
\usepackage{multirow}
\usepackage{geometry}
\newgeometry{
  top=72pt,
  bottom = 54pt,
  left = 54pt,
  right = 54pt
}

\usepackage{algorithm}
\usepackage{algorithmic}
\usepackage{array}
\usepackage{textcomp}
\usepackage{stfloats}
\usepackage{url}
\usepackage{diagbox}
\usepackage{verbatim}
\usepackage{graphicx}
\usepackage{multirow}
\usepackage{cite}
\usepackage{xcolor}
\usepackage{amsthm}

\newcommand{\Pref}{P^{\text{ref}}}
\newcommand{\fclki}{f_{\text{clk},i}}

\usepackage{stfloats}

\usepackage{tikz}
\usetikzlibrary{positioning,arrows.meta} 
\usepackage{graphicx}

\usepackage{bbm}

\def\BibTeX{{\rm B\kern-.05em{\sc i\kern-.025em b}\kern-.08em
    T\kern-.1667em\lower.7ex\hbox{E}\kern-.125emX}}

\usepackage{microtype}
\usepackage{float}
    
\begin{document}

\title{Voltage Regulation in Distribution Systems with Data Center Loads
\thanks{Y. Chen is partially supported by NSERC Discovery Grant RGPIN-2024-05015.  B. Zhang is partially supported by the University of Washington Clean Energy Institute, the Grainger Foundation and the Galloway Foundation.}
}

\author{
\IEEEauthorblockN{Yize Chen}
\IEEEauthorblockA{\textit{Department of Electrical and Computer Engineering} \\
\textit{University of Alberta}\\
Edmonton, Alberta, Canada \\
yize.chen@ualberta.ca}
\and
\IEEEauthorblockN{Baosen Zhang}
\IEEEauthorblockA{\textit{Department of Electrical and Computer Engineering} \\
\textit{University of Washington}\\
Seattle, Washington, USA \\
zhangbao@uw.edu\vspace{-15pt}}}

\maketitle
\begin{abstract}
Recent boom in foundation models and AI computing have raised growing concerns on the power and energy trajectories of large-scale data centers. This paper focuses on the voltage issues caused by volatile and intensity of data center power demand, which also aligns with recent observations of more frequent voltage disturbances in power grids. To address these data center integration challenges, we propose a dynamic voltage control scheme by harnessing data center's load regulation capabilities. By taking local voltage measurements and adjusting power injections at each data center buses through the dynamic voltage and frequency scaling (DVFS) scheme, we are able to maintain safe voltage magnitude in a distributed fashion with higher data center computing load. Simulations using real large language model (LLM) inference load validate the effectiveness of our proposed mechanism. Both the LLM power data and proposed control scheme are open sourced\footnote{\url{https://github.com/chennnnnyize/voltage-regulation-with-data-centers}}.

\end{abstract}

\begin{IEEEkeywords}
Voltage regulation,  AI, Data Center Energy Consumption
\end{IEEEkeywords}




\section{Introduction}

Large-scale data centers are in the midst of a period of explosive growth driven by advances in AI and compute. Applications such as large language models (LLMs) and other foundational models such as GPT-4, Llama 3, and BERT have pushed the boundaries of modern AI while consuming an increasingly significant amount of power and energy~\cite{bommasani2021opportunities, bubeck2023sparks, dubey2024llama}. The International Energy Agency (IEA) estimated that data centers consumed 460 TWh in 2022, accounting for 2\% of global electricity consumption--a figure that is still rapidly increasing and expected to reach $5\%$ by 2030 \cite{iea2024electricity}. Such energy burdens are generally observed in both the training and deployment stages of modern AI models \cite{chien2023reducing}, and are further exacerbated by the growing cooling needs for more advanced GPU clusters~\cite{nagendra2024thermal,zhu20241500}.

Much attention has recently been paid to both the average and peak power demand of data centers. And for good reason. As electric power makes up more than 60\% of the total operational cost of data centers~\cite{danfoss}, they face the complex task of meeting customers' quality of service requirements while balancing cost and sustainability. Meanwhile, power providers such as regional transmission operations and utility companies are having a challenging time coping with the ever-increasing and fluctuating data center demand. For example, despite skyrocketing electricity rates, California already has 270 data centers, and PG\&E customers are planning 24 new large-scale data centers, with an average load forecast of 3.5 GW, equivalent to the power consumption of almost 3 million residential households~\cite{pge2025datacenter}. Consequently, a number of works have proposed to reduce the total or peak power demand of data centers~\cite{liu2011greening,buyya2018manifesto,chien2023reducing}. 

In this paper, we do not directly address either of these issues. Instead, we focus on voltage regulations in grids with data centers, an issue we believe to be critical to their integration into the power system but has perhaps received less attention. Data centers, as well as many other devices, are sensitive to the voltage magnitudes of their grid connections. A short voltage disturbance in Northern Virginia in July 2024 led to more than 60 hyperscale data centers disconnecting from the grid and almost caused widespread blackout in the PJM system~\cite{Gooding25}.  A review by NERC also highlighted the challenges with voltage in data center load management~\cite{nerc2025}.

Voltage regulation itself is a classical problem in power system operations. Most distribution systems have several types of equipment dedicated to voltage support, including tap-changing transformers, switched capacitors and STATCOMs. In systems with inverter-based resources, many algorithms have been proposed to use the power electronics in the inverters to inject reactive power for voltage support~(see, e.g.~\cite{yeh2012adaptive,zhang2014optimal,srivastava2023voltage} and the references therein). 

\begin{figure}[ht]
    \centering
    \includegraphics[width=\linewidth]{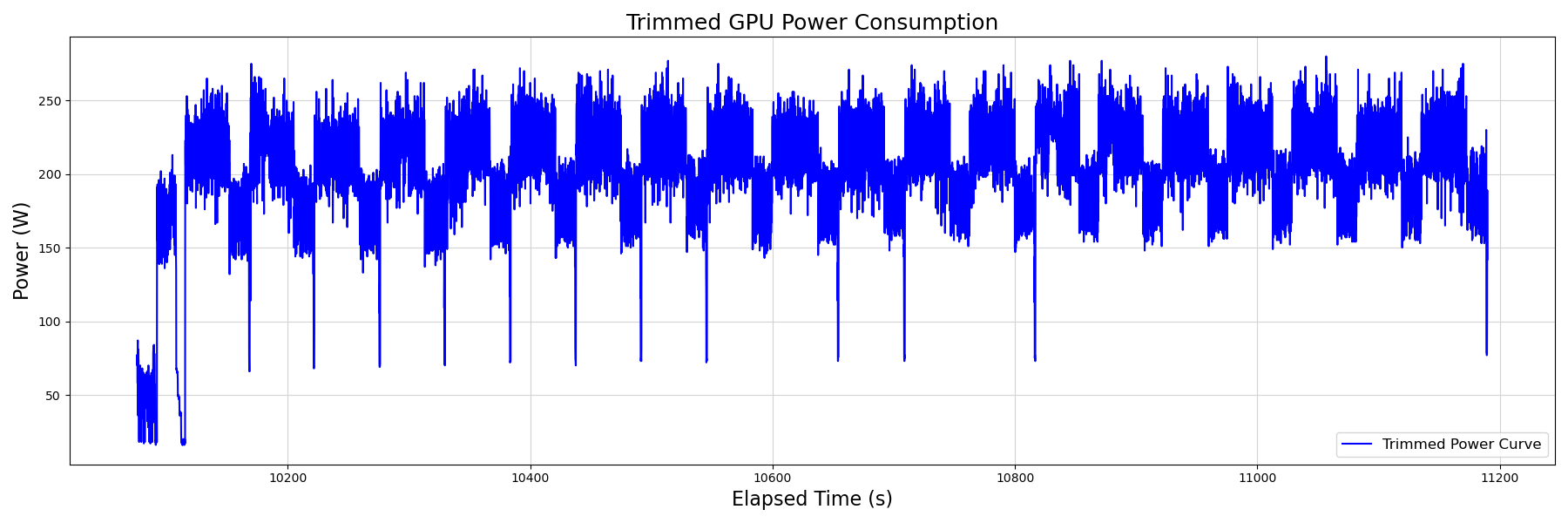}
    \caption{Power consumption curve for one GPU during the model training of GPT2 124M. Multiple GPUs are ramping up and ramping down almost simultaneously, creating highly volatile and huge disturbances to distribution grid's active power demand.}
    \label{fig:training}
\end{figure}
\begin{figure}[ht]
    \centering
    \includegraphics[width=\linewidth]{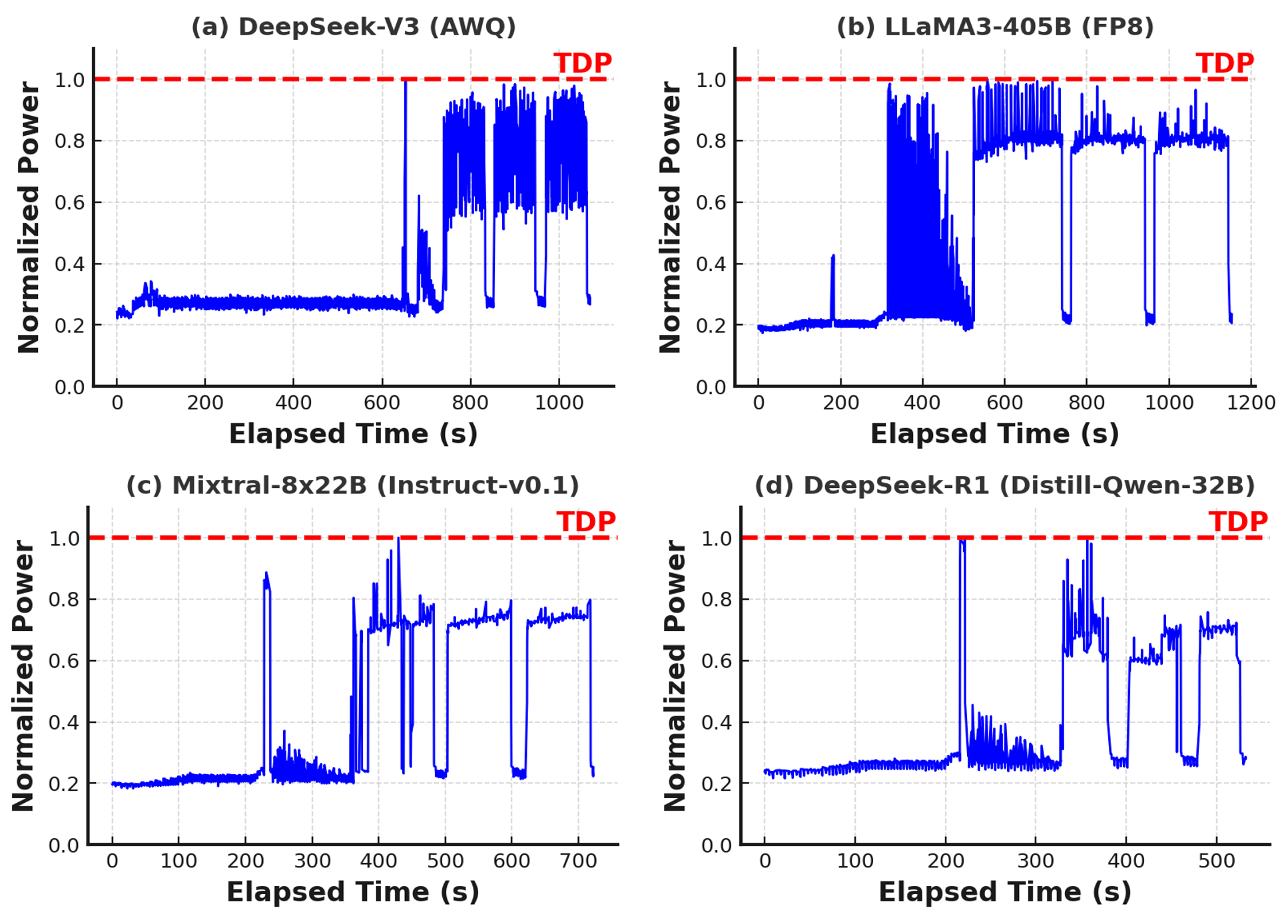}
    \caption{Power profiles of the four LLM inference tasks reveal distinct power ramping characteristics. Power is normalized in $[0, 1]$ range with thermal design power (TDP) limit marked.}
    \label{fig:inference}
\end{figure}
What makes data centers challenging and interesting in the context of voltage regulation is their unique load characteristics especially in the era of LLMs and foundational models~\cite{li2024unseen}. Fig.~\ref{fig:training} shows the single GPU's power consumption profiles of training a GPT2 124M model on a cluster consisting of Nvidia A100 GPUs;
each subfigure in Fig.~\ref{fig:inference} shows the power consumption profile of performing inference tasks using several common foundational models, and these inference timeseries is collected on a 8×A100 cluster instance. Also note that GPU power can be slightly over the Thermal design power (TDP) when GPU frequency is higher.

We observe sharp dips in power consumption during both inference and training.\footnote{These dips are much more important in transient dynamics compared to more ``steady-state'' type of analysis. They are too short to have much impact on total energy consumption, and because they represent a decrease in power draw, they would not factor into peak demand at all.} In these dips, a large down ramp is quickly followed (sometimes within a second) by a large up ramp, which can lead to large voltage deviations in the distribution system if left uncompensated. Because of the suddenness of the load changes, it is difficult to compensate for them using conventional devices such as tap-changing transformers or switched capacitors, which cannot act quickly enough to deal with (sub)second-level transients~\cite{yeh2012adaptive}.  In addition, large data centers are often placed at the end of large lines, and since reactive power injections tend to be "localized", power electronic devices placed away from the data centers would have limited impact, despite their fast reaction capabilities.  

Therefore, in this paper, we analyze how data centers could adjust their active and reactive power injections to regulate voltage. For reactive power compensation, we adopt the standard technique in which we utilize power electronic rectifiers to change the amount of reactive power injected into the grid. To change the active power, we use dynamic voltage and frequency scaling (DVFS) to control the dynamic power draw of the GPUs. More specifically, we focus mostly on changing the frequency of the GPUs~\cite{bao2016static,wang2021dynamic}. For example, NVIDIA GA100 supports operations at different frequencies in the range between 210 MHz and 1410 MHz~\cite{luo2024benchmarking}. Leveraging DVFS to adjust the underlying GPU’s frequency and power consumption in turn, it is possible achieve optimization at the iteration-level (spanning only milliseconds) opposed to the overall inference times (spanning several seconds). Since the power draw scales linearly with switching frequencies~\cite{agarwal2005foundations}, we can tradeoff compute with power. The circuit-level voltage (e.g., $v_{dd}$) can also be changed to manage power consumption, but it appears to be more difficult to access, and we do not consider it in this paper. Interestingly, DVFS is a common technique used in circuit design to manage power at the chip level~\cite{rabaey2002digital}, but it has not been leveraged to support the operation at the grid level.

\begin{figure}[h]
    \centering
    \includegraphics[width=0.8\linewidth]{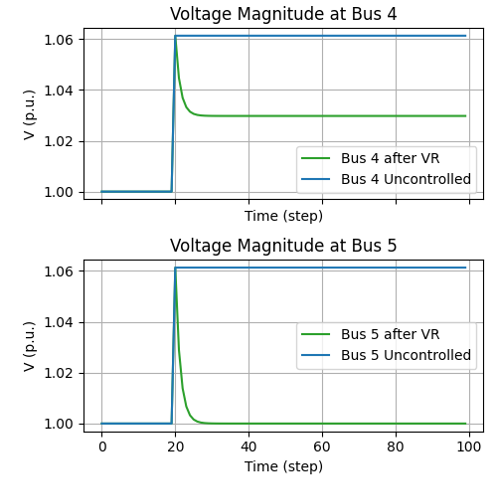}
    \caption{Voltages in a 6-bus toy example. Without voltage regulation, the voltage can float high after a change in compute load.}
    \label{fig:6bus}
\end{figure}
Active power management has been studied in the context of balancing renewables, reducing carbon emission and minimizing cost subject to some price signal~(see e.g., \cite{ORO2015429, KWON2020115424, kong2014survey,stojkovic2024dynamollm} and the references therein). However, these problems operate on slower timescales than voltage regulation. In addition, many distribution systems do not yet have real-time communication capabilities. Hence, the controllers in this paper have the flavor of primary controllers in power systems, where they take local measurements and adjust power injections at their own buses. We present the models in Section~\ref{sec:model}. We show how DVFS can be adapted to form an active power control loop (Section~\ref{sec:control}) that is quite effective in regulating the voltage as shown by the example in Fig.~\ref{fig:6bus} and larger simulations in Section~\ref{sec:simulations}.

\section{Modeling for Data Centers and Power Grids} \label{sec:model}
\subsection{Distribution Grid Voltage Regulation}

A standard requirement for distribution networks is that voltages should not deviate too far from their rated value, for example, not more than 5\% at all buses~\cite{national1996american}. For simplicity, we normalize the units such that the reference value for voltage is 1 p.u. at all buses. 
We consider a power network with $N+1$ buses, numbered as $0,1,\dots,N$, with bus $0$ as the feeder. Let $\bm{v}$ be the voltage vector where $v_i$ is the voltage at bus $i$. Let $\bm{p}$ be active power and $\bm{q}$ be reactive power, respectively. The voltage of the system follows the LinDistFlow model~\cite{zhou2020note}: 
\begin{equation}\label{eq:Dyn_voltage}
\bm{v}=\mathbf{R} \bm{p}+\mathbf{X} \bm{q}+\mathbbm{1};
\end{equation}
where $\mathbbm{1}$ is the all one's vector and $\mathbf{R}$ and $\mathbf{X}$ are positive definite matrices describing the network~\cite{zhu2015fast}. 

We assume some buses are data centers, some buses have inverter-based resources (e.g., solar PV or battery storage) and the rest of the buses are PQ loads. Our goal is to use the flexibility in data centers and inverter-based resources to control the voltages to be close to 1 p.u., at the same time minimizing the cost of using these resources. Since the control of inverter-based resources is not the focus of this paper, we assume that they run one of the numerous algorithms (for example, droop control on reactive power~\cite{cui2022decentralized}) that have been proposed and would otherwise not deal with their behavior until the simulation results.

\subsection{Control of AI Data Center Loads}
We now turn to the control of data center loads. We adopt a two axes control strategy, with control loops on both active and reactive power. Compared to other loads, data centers with AI loads tend to be very power intensive and operate 24/7, with sudden and quick fluctuations interspersed in between. Active power being at or near the rated capacity constrains reactive power capability of the power electronic converters~\cite{turitsyn2011options,streitmatter2025geometry}. In addition, because of the high $r/x$ ratios of power lines in distribution systems, voltage is some times more sensitive to changes in active power and hence it is more effective to do two axes control than just changing reactive power.

Since we do not assume the availability of real-time communication, $\bm{p}$ and $\bm{q}$ need to be successively updated based on the local voltage measurements. We consider a discrete time system indexed by time $t$. The desired active power consumption of the data center at bus $i$ and time $t$ is denoted as $\Pref_{i,t}$. We assume that $\Pref_{i,t+1}$ is predictable at time $t$, that is, we have full knowledge of the data center load in the next time step.\footnote{This is a reasonable assumption since the compute load of applications in data centers are predictable over a short timescale.} After observing its voltage $V_{i,t}$ at time $t$, bus $i$ sets its active power consumption in the next time step as $P_{i,t+1}=u^P_{i,t}(V_{i,t},\Pref_{i,t})$. Similarly, the reactive power at time $t+1$ is $Q_{i,t+1}=u^{Q}_{i,t}(V_{i,t})$, except there is not a desired value for reactive power. The system dynamics becomes
\begin{equation}\label{eq:Dynamic}
 \bm{v}_{t+1} =\mathbf{R} \bm{u}^P_t+\mathbf{X} \bm{u}^Q_t + \mathbbm{1}.
\end{equation}

We are interested in finding  $u^P_i$ and $u^Q_i$ for controllable bus $i$ that regulate the voltage close to $1$, satisfy their respective constraints and do not severely degrade the quality of service (QoS) of the data centers. We assume that we look at the problem over a total horizon of length $T$, with $T$ being a large number. A discounted infinite horizon formulation can also be used. We measure the voltage regulation performance at bus $i$ as the total deviation from the reference voltage $\sum_{t=1}^T |V_{i,t}-1|$. We require that the active and reactive powers are bounded, that is, $u^P_i \in [\underline{P}_i,\overline{P}_i]$ and $u^Q_i \in [\underline{Q}_i, \overline{Q}_i]$. To ensure the compute quality of service, we assume that the total power (or energy) consumed remains invariant, that is, $\sum_t P_{i,t}=\sum_t \Pref_{i,t}$ for a bus $i$ with a data center. 

Putting the dynamics together with the cost and the constraints, our goal is to solve the following optimization problem throughout all timesteps:
\begin{subequations} \label{eqn:opt_u}
\begin{align}
    \min_{\bm{u}^P, \bm{u}^Q} \; & \sum_{i,t} |V_{i,t}-1| \\
    \text{s.t. } &  P_{i,t+1}=u^P_{i,t}(V_{i,t},\Pref_{i,t}) \;\; \forall i\\
    & Q_{i,t+1}=u^{Q}_{i,t}(V_{i,t}) \;\; \forall i \\ 
    & \bm{v}_{t+1} =\mathbf{R} \bm{p}_{t+1}+\mathbf{X} \bm{q}_{t+1} + \mathbbm{1} \\
    & u^P_i \in [\underline{P}_i,\overline{P}_i] \;\; \forall i \\
    & u^Q_i \in [\underline{Q}_i, \overline{Q}_i] \;\; \forall i \\
    & \sum_t P_{i,t}=\sum_t \Pref_{i,t} \;  \text{, $i$ is a data center bus.} \label{eqn:power_balance}
\end{align}
\end{subequations}

Fig. \ref{fig:6bus} demonstrates a toy example on the effect of sudden load change on the voltage profiles before and after adding controls $\bm{u}$ in IEEE 6-bus system.



\section{Data Center Load Control} \label{sec:control}
In problem in \eqref{eqn:opt_u} is over all local controllers and solving it to optimality is challenging. We adopt a simple strategy based on linear droop control that we will show to perform quite well in simulations. 

The control knob we tune to control active power is the DVFS capability on the GPU chips. We assume that the GPUs within a data center are running with a common clock, which is the case when a single compute intensive AI application is being served.  Let $\fclki$ denote the clock frequency and let $\tilde{P}_i$ denote the power consumed at a data center without any other constraints or controls. The GPU power is then linearly related to the frequency:
\begin{equation}\label{eqn:p_f}
\tilde{P}_i \sim \fclki.
\end{equation}
The constant of proportionality can depend on a number of factors, including the exact hardware setup and the compute application. We control in turn control set the clock frequency as a linear function of the voltage deviation $\fclki \sim (V_{i,t}-1)$. Together, we obtain an active power control droop in the form of $k_{p,i} (V_{i,t}-1)$, with $\overline{P}_i$ and $\underline{P}_i$ determined by GPU $i$'s highest and lowest frequency available. 

Because of the total power balance constraint in \eqref{eqn:power_balance}, we cannot just set $u^P_i$ based on the voltage/power droop. Therefore we introduce a term $S_{i,t}$ that keeps track of the power not served up to time $t$. All together, active power control on bus $i$ is 
\begin{equation} \label{eqn:u_p}
u^P_{i+1,t} = [\Pref{i,t} - k_{p,i} (V_{i,t}-1) + \alpha_{i,t} S_t]^{\overline{P}_i}_{\underline{P}_i}, 
\end{equation} 
where $[a]^{\overline{a}}_{\underline{a}}$ thresholds a variable to the upper or lower bound ($[a]^{\overline{a}}_{\underline{a}}= \max(\min(a,\overline{a}),\underline{a})$). 
And $S_{i,t}$ evolves as 
\begin{equation} \label{eqn:S}
S_{i,t+1} = \Pref_{i,t}-u^P_{i+1,t}.
\end{equation} 

The $\alpha_{i,t}$ parameter is tuned to encourage the satisfaction of the total power balance constraint. For example, if $S_{i,t}$ is large (compute load not served), then we can have $\alpha_{i,t}$ increased. Conversely, it is decreased when $S_{i,t}$ is small. 

The reactive power control loop is simpler and takes the standard form of 
\begin{equation} \label{eqn:u_Q}
u^Q_{i,t+1} = [u^Q_{i,t} - k_{q,i} (V_{i,t}-1)]^{\overline{Q}_i}_{\underline{Q}_i}. 
\end{equation}
For buses with inverters but not data centers, we assume that they do not perform any active power control and adopts a reactive power control loop in \eqref{eqn:u_Q}. 









\section{Numerical Simulations} \label{sec:simulations}
\subsection{Simulation Setup}

\begin{figure}[h]
    \centering
    \includegraphics[width=0.99\linewidth]{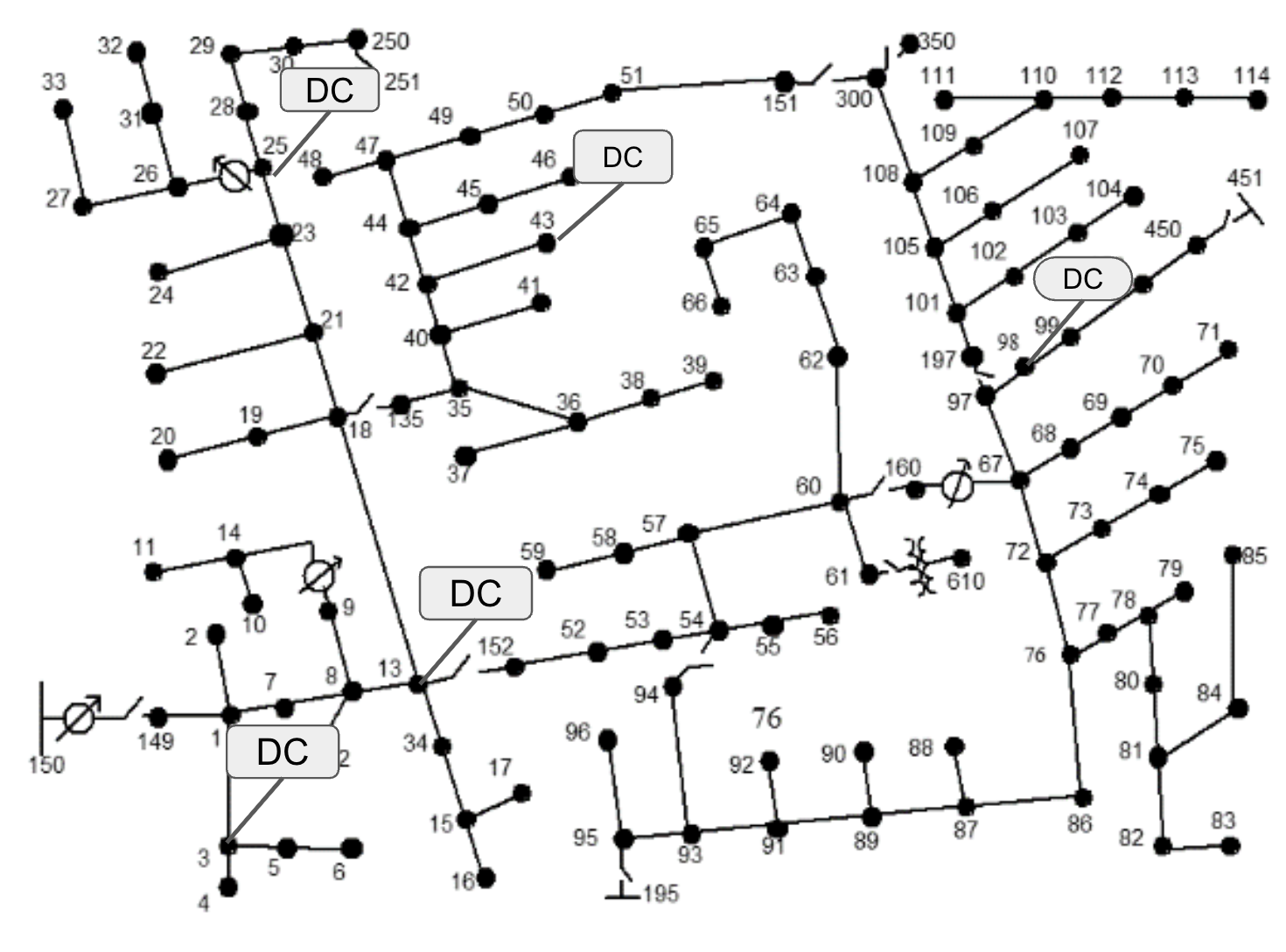}
    \caption{IEEE 123-bus system with AI data center load buses.}
    \label{fig:123-bus}
\end{figure}

\textbf{Data Center Load}: To validate our proposed approach is practical, we use real LLM inference load in the IEEE 123-bus test system (Fig.~\ref{fig:123-bus}), so that both the impacts of AI data center load and effectiveness of data center-enabled voltage regulation. The inference load presented in this study is conducted on an 8× A100 cluster instance. Each GPU offers 80GB of high-bandwidth memory, allowing simultaneous data-parallel and model-parallel strategies for LLM inference tasks. We use standard set of prompts and query the LLama 3.1 70B Instruct Model. In practice, we see high GPU utilization when the GPUs are loaded with LLM inference tasks. Since neural transformer-based LLM models are autoregressive which generate output on a token-by-token basis, they require multiple iterations and the inference duration are diverse and depending on the query length. Correspondingly we observe large and fast swings in power draw, particularly during steps involving heavy matrix multiplications and data communications. When the LLM query firstly arrives or inference task finishes, there is usually a huge ramping up or ramping down event in the power demand signal.

After collecting the power measurement curves under a variety of prompts, we scale the GPU load to mimic the load behavior of a small- to medium-sized data center with peak workloads up to $280\,kW$. We then randomly select snippets of such data center loads and assign them randomly to the data center nodes, which simulates the random arrival patterns for LLM tasks in AI data centers. Note that in the current simulation we practice first-come, first-serve (FCFS) scheduling for LLM queries, while analyzing impacts of batching and scheduling policies as well as congestion control are left as future work.

\textbf{Power Grid Simulations}: We modify the standard IEEE 123-bus by randomly selecting nodes to equip with inverters or data centers. For the inverter buses, we assign the same bounds for reactive power injections, that is, the inverters are sized uniformly. We keep the number of inverters and inverter buses fixed in the network, but assign 1 to 5 random data center buses to examine the effects of high penetration of data center loads (Fig. \ref{fig:123-bus}). For all the other load buses, we perturb the standard load profiles  provided in the 123-bus system with Gaussian noises. The active power is balanced by slack bus's active power injections at every timestep $t$. Throughout the simulations, we set the simulation interval and control interval as 1 second. While we note that the proposed droop-based control strategy adds very mild computation, and can be implemented in a faster timescale. 

In the simulations, for added accuracy, we actually compute $\bm{v}_{t+1}$ using the nonlinear distflow equations~\cite{baran2002optimal}. The results are not qualitatively different if linear equations are used. 



\begin{table*}[htbp]
\centering
\caption{Voltage Magnitude Deviation and LLM Query Delay with Varying Number of Data Center Nodes in 123-Bus System}
\begin{tabular}{ccc|cc|cc}
\toprule
\multirow{2}{*}{\textbf{No. of Data Centers}} & \multicolumn{2}{c}{\textbf{$k_p=1$}} & \multicolumn{2}{c}{\textbf{$k_p=10$}} & \multicolumn{2}{c}{\textbf{$k_p=20$}} \\
\cmidrule(lr){2-3} \cmidrule(lr){4-5} \cmidrule(lr){6-7}
& $\Delta V$ (p.u.) & Average Delay (s) & $\Delta V$ (p.u.) & Average Delay (s) & $\Delta V$ (p.u.) & Average Delay (s) \\
\midrule
1 &$0.020\pm 0.012$  & $0.3$ & $0.020\pm 0.013$ & $6.0$ & $0.019\pm 0.012$ &  $11.2$\\
2 &$0.021\pm 0.012$  & $0.5$ & $0.021\pm 0.013$ & $5.6$ & $0.020\pm 0.011$ &  $9.2$  \\
3 &$0.026\pm 0.013$  & $0.5$ & $0.024\pm 0.013$ & $6.2$ & $0.022\pm 0.013$ &  $13.7$  \\
4 &$0.027\pm 0.015$  & $1.1$ & $0.023\pm 0.013$ & $6.0$ & $0.021\pm 0.015$ &  $13.8$  \\
5 &$0.028\pm 0.017$  & $1.4$ & $0.024\pm 0.016$ & $6.0$ & $0.024\pm 0.016$ &  $17.5$  \\
\bottomrule
\label{table:control}
\end{tabular}
\end{table*}

\subsection{Simulation Results}
\begin{figure}[htp]
    \centering
    \includegraphics[width=0.9\linewidth]{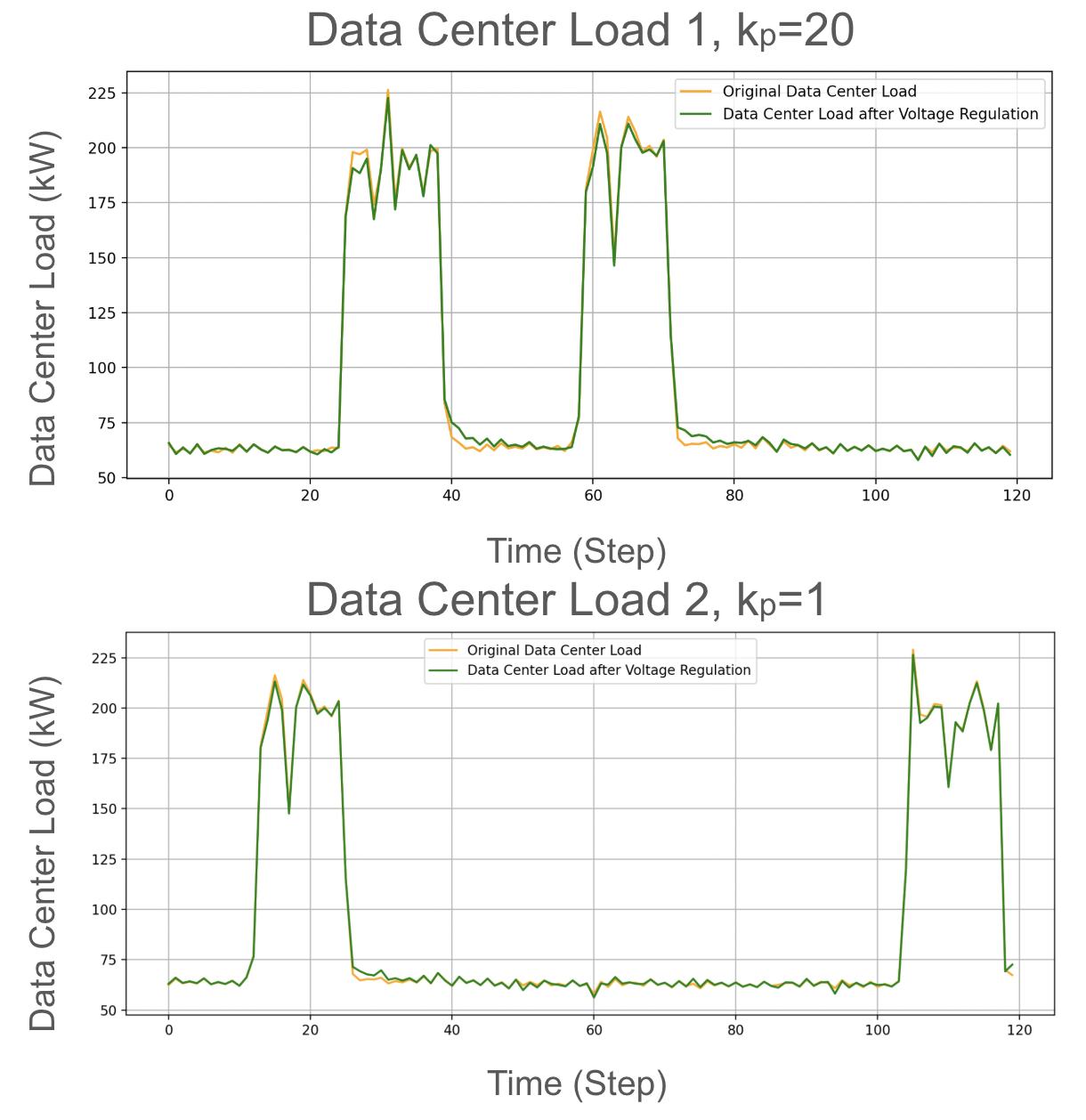}
    \caption{Data center original load curve and load curve after implementing voltage regulation under $k_p=20$ and $k_p=1$.}
    \label{fig:load_curve}
\end{figure}

It can be revealed in Fig. \ref{fig:training}, Fig. \ref{fig:inference}, and Fig. \ref{fig:load_curve} that to accommodate large number of LLM weights and matrix operations, both LLM training and inference tasks exhibit large power ramps. Due to the nature of autoregressive model, the valley-peak ramping behaviors exhibit longer periodic patterns, while the query length and thus peak load time is highly diverse. During inference stages, larger LLMs such as Llama3 405B do not always exhibit more dynamic power behavior—mid-scale models like Mixtral-8x22B and DeepSeek-V3-AWQ, which show more pronounced fluctuations. The 10–20\% ramping range dominates, indicating frequent computations tied to attention mechanisms inside LLMs. 

\begin{figure}[h]
    \centering
    \includegraphics[width=0.8\linewidth]{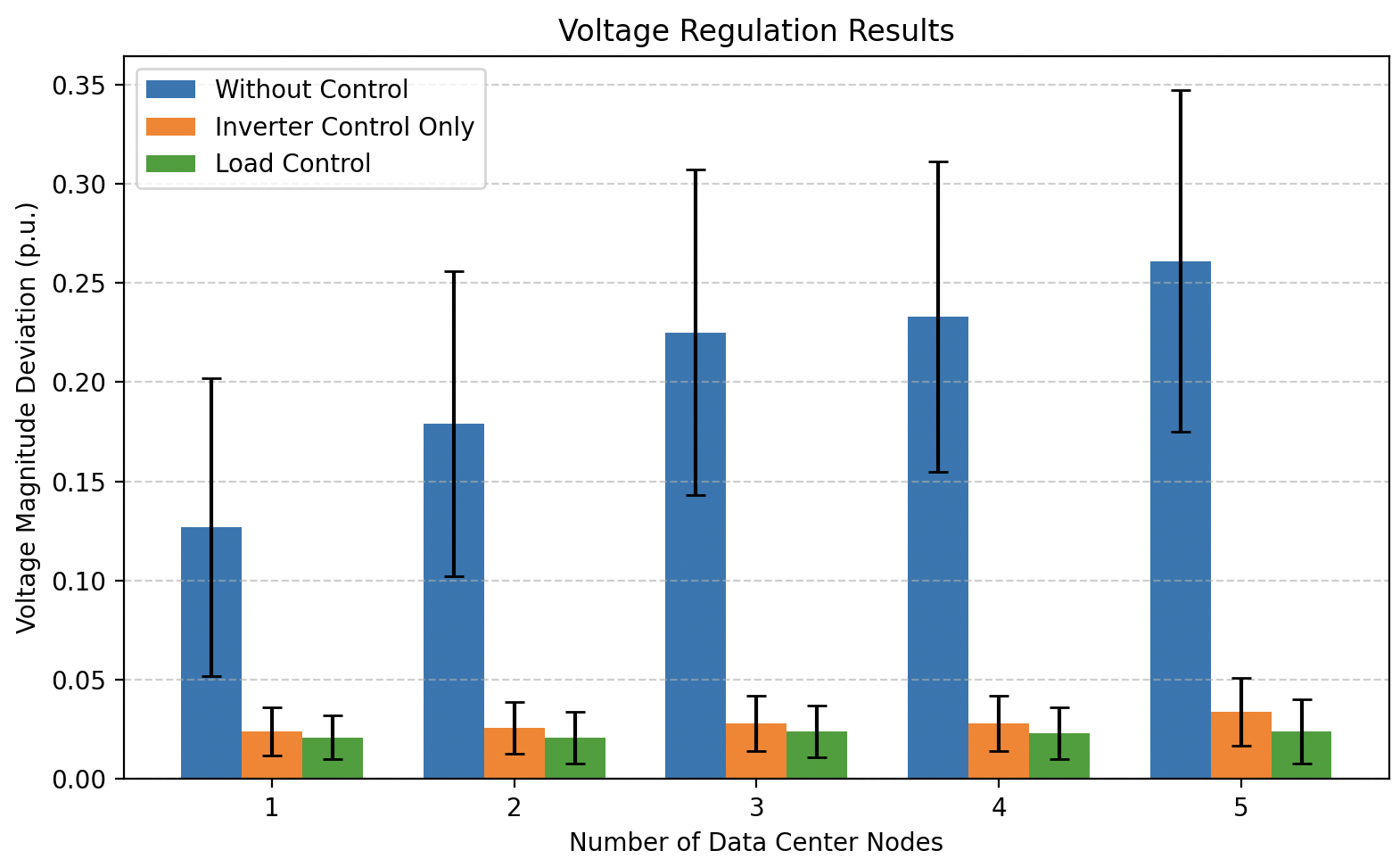}
    \caption{Average voltage magnitude deviation (p.u.) and standard deviation with varying number of AI data center load buses.}
    \label{fig:voltage}
\end{figure}

Fig. \ref{fig:voltage} visualizes the voltage deviation from the nominal values under three settings: plain data center integrations without inverter nor data center load control. It can be clearly observed that without inverter or data center voltage control, the added data center load will heavily affect the voltage profile, and a large voltage deviation is observed. With growing number of data centers from 1 to 5, average voltage deviation tends to increase for all three settings. While it is noteworthy by integrating data center load control, the least deviation in voltage profile is recorded. More than $97\%$ of the buses always have a voltage deviation within $0.05 \, p.u.$ throughout the simulation horizon. On average, such a control scheme achieves more than $12.8\%$ gain compared to inverter-based droop control only in terms of voltage magnitude deviation.

\begin{figure}[h]
    \centering
    \includegraphics[width=0.9\linewidth]{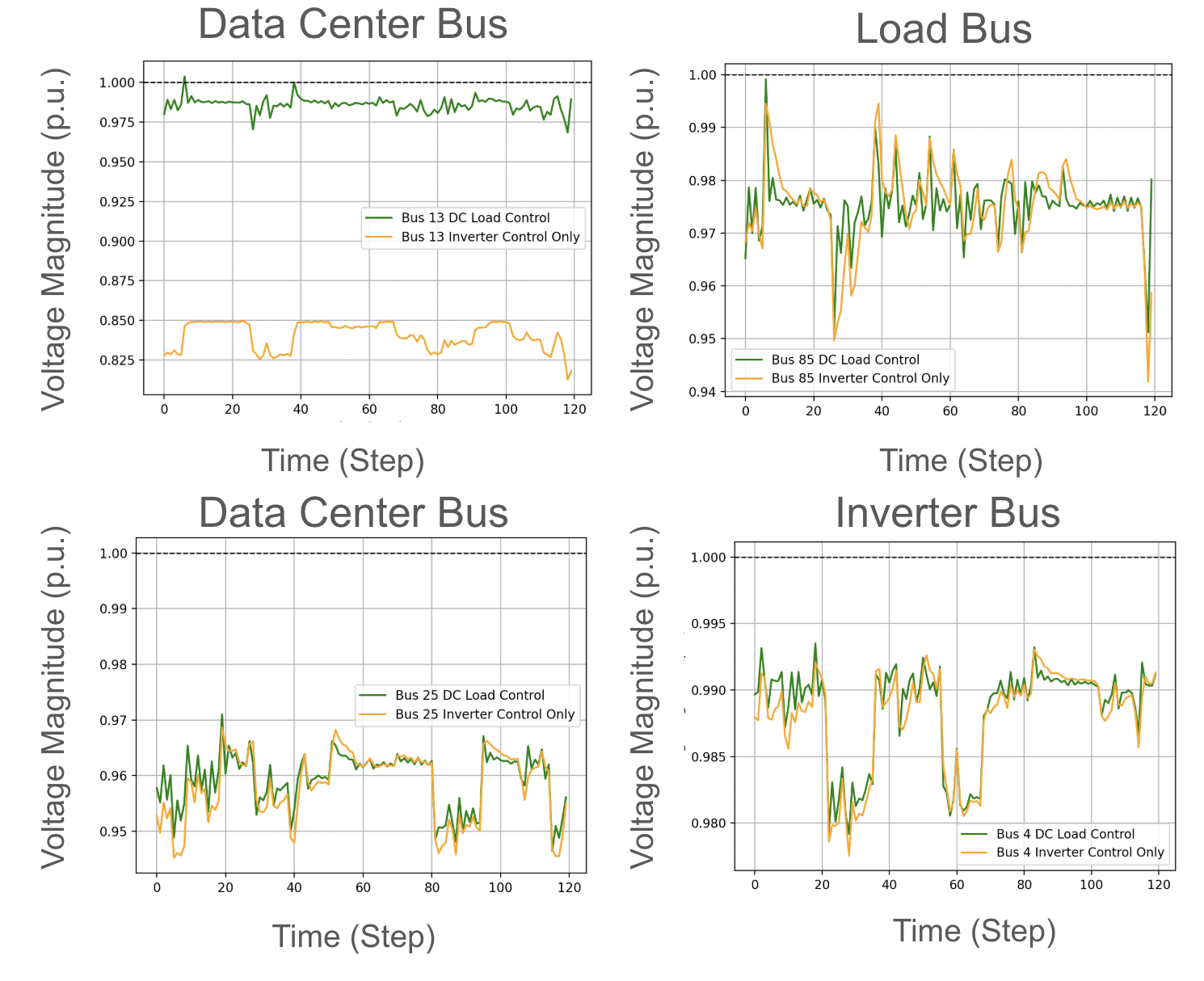}
    \caption{2-minute voltage curves for selected two data center buses, a nominal bus, and an inverter bus. Control involving only inverters and involving data centers with $k_p=10$ are compared.}
    \label{fig:voltage_curve}
\end{figure}

We evaluate the voltage regulation effects with varying droop parameters $k_p$. The results are summarized in Table. \ref{table:control}. For all choices of $k_p$ and varying degree of data center integrations (number of data centers from 1 to 5), the buses' average voltage deviation can be controlled within $0.03\, p.u.$. There is a mild increase in voltage deviation when the number of data center grows, as it becomes more challenging to regulate more volatile load. With $k_p=10$ and with $5$ data center load buses, proposed coordinated control of data centers and inverters can achieve an average voltage magnitude deviation of $0.024$, while with only inverter control, such deviation can reach  $0.034$. Moreover, as shown in Fig. \ref{fig:voltage_curve}, bus 13's voltage is out of the $[0.95, 1.05]$ p.u. range with only inverter control. This is partly because bus 13 has 4 adjacent connected neighbors, making the voltage more sensitive compared to other buses. In addition, the larger data center load at bus 13 also makes it challenging to control the voltage. Further, as illustrated in Fig. \ref{fig:voltage_curve}, with data center actively participating the voltage regulation, voltage magnitude for all buses are closer to $1.0$ p.u..

For each data center, the $k_p$ parameter also decides the tradeoff between load flexibility and voltage regulation participation. As indicated in  Table. \ref{table:control}, when $k_p=20$, the LLM inference query experiences a signficant delay compared to smaller $k_p$. This is because when there is a ramping up in active load curve, the droop control strategy implicitly reduce such fast load increases, and shift them to latter intervals. Similarly, original LLM tasks are having sudden endings when the end-of-generation tokens are generated. While under droop control, the data center gradually decreases load. Thus, in practice, striking a balance in the choice of the $k_p$ parameter is essential for optimizing the performance of data centers under dynamic load and voltage conditions.

In Fig. \ref{fig:load_curve}, we visualize two randomly selected data center load curves before and after implementing voltage regulation on data center buses. Compared to the original load, we can observe that implementing active load control via DVFS will smooth the load curve, and cause a mild delay in completing the LLM token generation tasks.

\section{Conclusion}
In this work, we look into the voltage issues with growing data centers connecting to the grid. We first identify the voltage regulation challenges caused by sheer volume and stochasticity of AI data center's power demand. To effectively regulate distribution grid voltage, we show it is possible to convert data center as flexible load resources by utilizing GPU's DVFS capabilities. By leveraging a distributed droop control scheme, we propose a simple and effective linear control strategy that dynamically adjust LLM workload as well as inverter buses' injections. Simulation results on real LLM inference load validate the effectiveness of our proposed approach, demonstrating significant improvements in grid voltage profiles while balancing LLM query delays. In the future work, we will investigate more fine-grained control knobs available in GPUs and AI data centers, and explore the integration of these computing-side control mechanisms with existing grid-side management systems.

\bibliographystyle{IEEEtran}
\bibliography{references}



\end{document}